# A Game-theoretic model of forex trading with stochastic strategies and information asymmetry

Patrick Naivasha, George Musumba, Patrick Gikunda, John Wandeto

**Abstract**

Interaction strategies for reward in competitive environments are significantly influenced by the nature and extent of available information. In financial markets, particularly foreign exchange (forex), *traders* operate independently with limited information, often yielding highly unpredictable outcomes. This study introduces a game-theoretic framework modeling the market as a strategically active participant, rather than a neutral entity, within a stochastic, imperfect-information setting. In this model, the market alternates sequentially with new *traders*, each *trader* having limited visibility of the market's moves, while the market observes and counteracts each trader's strategy. Through a series of simulations, we show that this information asymmetry enables the market to consistently outperform *traders* on aggregate. This outcome suggests that real-world forex environments may inherently favor market structures with greater informational advantage, challenging the perception of a level playing field. The model provides a basis for simulating skewed information environments, highlighting how strategic imbalances contribute to *trader* losses. Further optimization of the *intelligent market's* scoring and refined simulations of trader-market interactions can enhance predictive analytics for forex, offering a robust tool for market behavior analysis.

## 1 Introduction

Speculating either a rise or fall of the exchange rate may be trivialized as a simple binary problem but the observation by [1] that only 2% of retail traders can successfully predict currency movement in the forex market is an indicator that the task is fairly complex for many. Interaction data generated during forex trading which is described as nonlinear, chaotic, extremely noisy, and non-stationary confirms this complexity [2]. Data protection regulations [3,4] hinder access to entire forex trading data which inhibits analysis. Modeling forex market presents an opportunity to benefit from some of the advantages identified by [4] such as addressing data privacy concerns and generation of rare data. This study introduces a game-theoretic framework modelling the market as a strategically active participant, rather than a neutral entity, within a stochastic, imperfect-information setting. The players are non-cooperative and market entry is sequential. An imperfect information game is a game theoretic construct in which players do not have complete knowledge of all elements influencing the game state [5]. *Traders* or market participants lack full visibility into market drivers or the intentions of competitors and is typical of our game model which we anticipate will provide critical insights into strategic behaviour under uncertainty. A non-cooperative game is characterized by the absence of binding agreements or collaboration among players, as each player acts solely to maximize their individual payoff, independently of the objectives of others. Within non-cooperative frameworks, individual optimization often leads to complex strategic interdependencies [6] which is encountered in forex trading and is replicated in our game model. A sequential game represents a structured framework in which players make decisions in a designated order, with each player's choice contingent upon the previous decisions of others. This sequential structure introduces a temporal dimension to strategy, allowing participants to adapt their actions based on observed moves, anticipated shifts, or responses from earlier stages [7]. This structure enables *intelligent market* to counter each *trader's* decision in real-time, strategically leveraging its responses to incrementally shift outcomes to its advantage. By capturing the temporal dependencies and adaptive strategies, this sequential framework provides a realistic basis for analysing market dynamics where systematic responses can shape cumulative outcomes.

*1.1 Objectives*

1) Model a market to strategically outperform *traders* by leveraging information asymmetries and varying levels of observational access
2) Analyze characteristics of the resulting interaction data to determine resemblance with real forex data

*1.2 Theoretical framework for strategic decision-making and Learning in competitive environments*

Game theory and reinforcement learning offer powerful frameworks for modeling complex real-world scenarios involving strategic interactions among participants [8,9], though their terminologies often differ. For example, the term "agent" in reinforcement learning corresponds to "player" in game theory. In this study, we employ terminology from both fields interchangeably to capture the cross-disciplinary nature of our approach. Typically, game models are represented through tables (normal-form) or trees (extensive-form), which delineate agents' actions, available information, associated values, and the outcomes of various strategic decisions [10]. The game model developed in this study has two primary players engaging in turns and denoted as *trader$_i$* and *intelligent market*, who alternate actions to simulate a predictive market scenario.

In order to inflict loss to every new player immediately after market entry, minimax strategy has been utilized. Minimax strategy is a decision-making approach widely used in two-player zero-sum games where maximizing player (*max*) tries to maximize their score or minimize the opponent's advantage while the minimizing player (*min*) tries to minimize the maximizing player's score or maximize their own advantage [11]. It is formulated as:

$$x \in X, y \in Y \max\min f(x,y) = y \in Y, x \in X \min\max f(x,y) = v \qquad (1)$$

*Where maxmin is interpreted as Player 1 choosing a strategy that maximizes their minimum payoff, minmax is interpreted as Player 2 choosing a strategy that minimizes Player 1's maximum payoff while v is the equilibrium value of the game*

The *intelligent market* is the *max* player and the strategy is actualized every time a player entering the game makes a move. New trader plays *min* if choice of price direction is the one on aggregate having the positive score (or lesser negative score). A Nash equilibrium condition can arise due to intelligent market elevated view but our game formulation does not exploit it.

*1.3 Drawing parallel of our Game model with forex market*

We have reviewed our game model and forex market in light of [12] description of cooperative and non-cooperative games. In their study, cooperative games are concerned with the participants forming alliances and working together seeking to achieve common goals while non-cooperative games are considered suitable for analyzing the behaviors of players who believe that their own payoffs conflict with the payoffs of others in the game. [13] observes that cooperation between agents for mutual benefits may arise even without aligned incentives. It is further indicated from the same study that for this to occur, cooperative intelligence which is comprised of understanding, communication, commitment and social infrastructure between the players is a necessity. In our game model, the game is non-cooperative because none of these requirements could be satisfied in the interaction environment. Studies by [14] indicate that there is no credible evidence that rewards for retail forex *traders* when viewed as a group are better than random trading. This observation emphasizes our model view and we have demonstrated analytically that skill and knowledge of *traders* might not influence their reward when playing against a market with intelligence.

Forex trading, as noted by [15], naturally entails transaction costs arising from broker fees and platform charges. Total game wealth in our framework is defined by the aggregation of *trader* gains, *trader* losses, and the *intelligent market's* accrued reward, which itself encompasses both declared and latent costs. These three principal components collectively establish the basis for our zero-sum calculations, mirroring the structural dynamics observed in forex trading environments.

While our model specifies that players operate independently without alliances, it is also crucial to establish that the design explicitly prevents collusion. [16] defines collusion as a covert agreement between parties that can undermine market efficiency, diminish consumer welfare, and hinder economic growth. As [17] noted, there remains no legal or documented evidence of algorithmic collusion in price-setting. In our model, although interactions among players are integral, they do not influence the price directly. The structure of the game inherently restricts any attempt at collusion or strategic manipulation, as outcomes are limited to binary price movements—up or down. Consequently, collusive strategies are rendered ineffective: if participants speculate in opposing directions, their rewards cancel out entirely or partially, and if they align, they face shared exposure rather than advantage.

*1.4 Integration of stochastic behaviour in the model*

Our game model has used random uniform distribution to label a trade as either a buy or sell and further used the same stochastic procedure to assign a value in the range between 0 and 1 to as the trade risk appetite. This introduces variability and unpredictability that reflects real-world trading behaviors consequently simulating the uncertainty *traders* face due to imperfect or incomplete information, which is a core characteristic of stochastic models [18].

*1.5 Forex trading reward structure*

Demonstrate that the reward is on the basis of magnitude of deviation and is proportionate to confidence value attached to predicted direction [19]. A general formula for analytical purposes would be formulate as:

*Reward= (magnitude of ratio deviation) \*(risk appetite) \*constant;* (2)

*where magnitude of ratio deviation may be positive if ratio deviates to predicted direction or negative otherwise while the constant is associated with currency pair being traded and the currency the trading account is denominated.*

To illustrate profit computation for hypothetical EURUSD currency pair trade, let a *trader* M with a trading account denominated in USD open a buy trade at a price of 1.0250 with a risk appetite of 0.01. If the trade is closed at 1.0350, accrued profit would be: (1.0350-1.0225) * 0.01 * 100,000 =$12.5. The constant 100,000 is the standard lot size in forex trading.

If we assume $trade_1$ entered the market at time $t_1$, and has a reward at $t_n$ of $trader_1\_reward$, then n *traders* reward at time $t_n$ would be:

$$\text{Aggregate traders reward} = \sum_{i=1}^{n}(trader_i reward) \quad (3)$$

Sustaining negative value for equation '3' implies a positive reward for *intelligent market* to satisfy zero-sum.

## 2 Methodology

Proceeding section outlines the setup of the model.

### 2.1 Player Description

*Traders*(buy/sell) can speculate a price will move up (open a buy) or down (open a sell) and associate the direction with a confidence level (risk appetite). *Intelligent market* adjusts ratio proportional to risk appetite against the choice of a *trader*.

### 2.2 Strategy

New *trader* and *intelligent market* alternately play the game. A *trader* predicts the direction the price will move then *intelligent market* counteracts by adjusting price in the opposite direction with magnitude proportionate to risk appetite. The initial price is pre-set.

Let:
- $r_0$ represent the initial price
- $r_t$ represent the price after t turns
- $s_i \in \{+1, -1\}$ be the speculation by $trader_i$; +1 predicts an increase, −1 predicts a decrease
- β be influence the *intelligent market* has on price adjustment where β= ($Trader_i$ risk appetite) * (0.001). Constant 0.001 is used to ensure resultant value is not a fractional pipette. In the context of the foreign exchange market, a pip is a standard unit of measure for changes in an exchange rate, representing a price move of 0.0001 (1/10,000). A pipette equals 1/10 of a pip and represents a move of 0.00001(1/100,000). This is the smallest price change increment for most currency pairs.

The price adjustment after $trader_i$ speculation can be represented as:

$$r_{(t+1)} = r_{(t)} + \beta \cdot s_i \quad (4)$$

After all n *traders* make their predictions and all counter moves by *intelligent market* the final price $r_f$ is:

$$r_f = r_0 + \sum_{i=1}^{n} \beta \cdot s_i \quad (5)$$

### 2.3 Reward Computation

The reward function for $trader_i$ is:

$$trader_i\_reward = \delta * (r_f - r_{ti}) * s_i \quad (6)$$

where:
- $r_f$ is the final price

- $r_{ti}$ is the price when trade was opened
- δ is a scaling factor which is the product of trader$_i$ risk appetite and a constant 100000 as used in equation '2'
- $s_i$ is the speculative direction chosen by trader$_i$; (either +1 or −1).

The aggregate reward for n *traders* is as per equation '7':

$$Aggregate\ traders'\ reward = \sum_{i=1}^{n}(trader_i reward) = \sum_{i=1}^{n} \delta * (r_f - r_{ti}) * s_i \qquad (7)$$

*Intelligent market* reward added to aggregate *traders*' reward results to predefined result of zero to align with our zero-sum model objective.

*2.4 Algorithm*

**Variables initialization:**
Set agg_traders_reward to 0.0 (accumulates aggregate reward for *traders*)
Set intelligent_market_reward to 0.0 (stores *intelligent market* reward)
Define function game_simulator (n_trades, initial_price):
- Initialize constant ppt to 0.001
- Set current_price to initial_price
- Initialize empty lists: price_pro (to store price progression), t_type (to store trade types), r_appetite (to store risk appetites), t_price (to store trade prices)

**Simulate trades:**
For each trade from 0 to n_trades - 1:
- Randomly generate a trade type ('0' for buy, '1' for sell)
- Generate a random risk appetite between 0 and 1
- Append trade type to t_type, and risk appetite to r_appetite
  **If the trade type is '0' (buy):**
    - Append current_price to t_price.
    - Calculate price_adjustment as ppt * risk_appetite.
    - Subtract price_adjustment from current_price to get new_price.
    - Append new_price to price_pro.
  **If the trade type is '1' (sell):**
    - Append current_price to t_price.
    - Calculate price_adjustment as ppt * risk_appetite.
    - Add price_adjustment to current_price to get new_price.
      - Append new_price to price_pro.
- Update current_price to new_price

**Return values:**
- After all trades, return the t_type, r_appetite, t_price, and price_pro.

**Run the simulation:**
- Call game_simulator with n_trades and an initial price of 1.0828.
- Store the returned trade types, risk appetites, trade prices, and price progression.

**Plot price progression:**
- Plot t_price against trade numbers

**Calculate *Traders* reward:**
- Loop through all n_trades trades:
    for i in range (n_trades):
        if t_type[i] == '0':
        agg_traders_reward+=(price_pro[n_trades-1]-t_price[i]) *100000*r_appetite[i]

```
    else:
        agg_traders_reward +=(t_price[i]-price_pro[n_trades-1]) *100000*r_appetite[i]S
```
**Calculate and display *Intelligent Market* reward://Ensure zero sum**
```
        intelligent_market_reward=agg_traders_reward * (-1)
```

*2.5 Evaluation*

Multiple batches of random risk appetites with their corresponding random trade types were generated and maximum price deviation on either side of open price for all intervals computed and compared with real data. *Intelligent market* reward is computed to satisfy the requirement of zero sum for all market participants. EURUSD has been adopted as the currency pair and reward computation assumes USD denominated *trader* account.

*2.6 Results*

### 2.6.1 Model Data

Table 1 illustrates price progression data of one of the simulations and corresponding line graph on Figure 1.

*Table 1: Trade attributes and ratio progression*

| Time | Trade Type | Risk Appetite | Price Progression |
|---|---|---|---|
| 1 | 0 | 0.43 | 1.08237 |
| 2 | 0 | 0.56 | 1.08181 |
| 3 | 0 | 0.27 | 1.08154 |
| . | . | . | . |
| . | . | . | . |
| . | . | . | . |
| 199999 | 1 | 0.9 | 1.08123 |
| 200000 | 0 | 0.41 | 1.08082 |

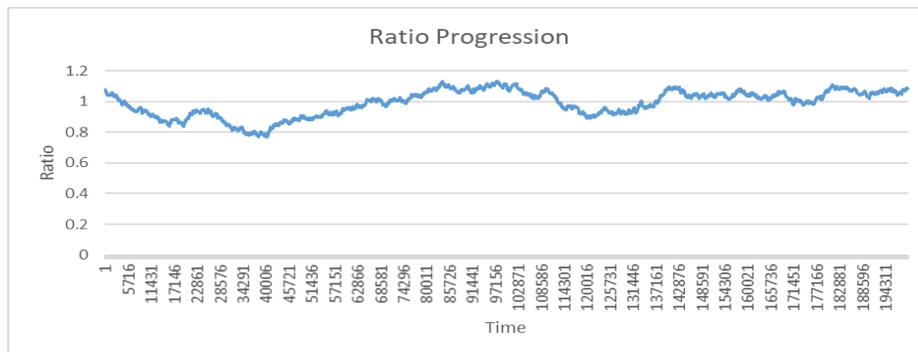

*Figure 1: Price against time line graph of table 1*

### 2.6.2 Intelligent market reward

Table 2 indicate the *intelligent market* reward for 10 independent simulations each with 200,000 data points of price

*Table 2: Intelligent market rewards for ten independent simulations*

| Simulation | 1 | 2 | 3 | 4 | 5 |
|---|---|---|---|---|---|
| Reward | 15905698.67 | 10076808.79 | 3331471.94 | 3367058.52 | 7757940.25 |

| Simulation | 6 | 7 | 8 | 9 | 10 |
|---|---|---|---|---|---|
| Reward | 4336847.49 | 6581004.17 | 12193555.46 | 10337426.06 | 8207841.06 |

### 2.6.3 Open, High, Low and Close (OHLC) data format

We have adopted one-second interval from one price to another in Table 1 so that our data can be presented as per the most granular interval provided by [20] i.e. 1-minute interval. Further, we have clustered the data in Table 3 into clusters of 60 consecutive prices (one hour) then extracted open price for the interval(Open), closing price (Close), highest deviation uptrend from open(High) and similarly for downtrend(Low). This is to conform to commonly requested fields of (Open, High, Low and Close) for forex studies as indicated by [20]. See Table 3 and Figure 2 pie chart comparing the cumulative total deviations for the intervals.

*Table 3: Hourly interval deviation for generated Data.*

*Open value is the price at the beginning of interval, close is the price at the end of that interval. High is the highest attained price (uptrend) within the interval while low is lowest attained price (downtrend) within the interval. +veDEV is highest magnitude of price attained in uptrend converted into pipettes while -veDEV is lowest magnitude of price attained in downtrend converted into pipettes*

| Hr | OPEN | HIGH | LOW | CLOSE | +veDEV ((High-Open)*100000) | -veDEV ((Open-Low)*100000) |
|---|---|---|---|---|---|---|
| 1 | 1.08237 | 1.08237 | 1.00398 | 1.01008 | 0 | 7839 |
| 2 | 1.01019 | 1.01379 | 0.94084 | 0.94454 | 360 | 6935 |
| 3 | 0.94406 | 0.96593 | 0.91568 | 0.92082 | 2187 | 2838 |
| . | . | . | . | . | . | . |
| . | . | . | . | . | . | . |
| . | . | . | . | . | . | . |
| 54 | 1.05926 | 1.08864 | 1.04354 | 1.07486 | 2938 | 1572 |
| 55 | 1.07413 | 1.09167 | 1.03819 | 1.05564 | 1754 | 3594 |

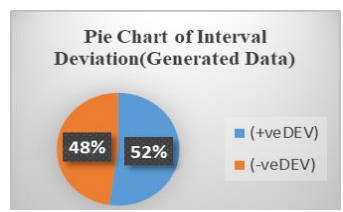

*Figure 2: Generated data hourly cumulative interval deviation.*
*Pie chart comparing +veDEV and -veDEV of Table 3*

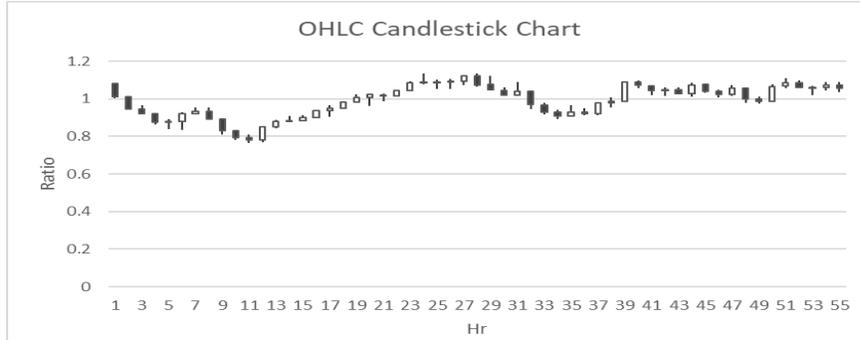

*Figure 3:* Candlestick chart for 55 Hourly intervals of Table 3

*Table 4:* Simulated data % cumulative deviation

| Simulation No. | 1 | 2 | 3 | 4 | 5 | 6 | 7 | 8 | 9 | 10 |
|---|---|---|---|---|---|---|---|---|---|---|
| % +ve deviation | 63 | 42 | 48 | 50 | 46 | 45 | 45 | 54 | 56 | 43 |
| % -ve deviation | 37 | 58 | 52 | 50 | 54 | 55 | 55 | 46 | 44 | 57 |

*2.7 Model evaluation*

Real EURUSD forex data was obtained from [21], a website that provides historical data for currency pairs among other forex trading resources. From the live stream, one-month historical data was captured for the period Sep 18th to Oct 29, 2024. Out of the 705 hourly intervals obtained, 10 batches of 55 consecutive intervals were extracted for comparison with our model data. See Table 5, Figure 4 and Figure 5.

**2.7.1 Interval deviation for generated data and real data**

Based on data in Table 7 and 8, maximum price deviations on either side from open price were computed for each of the intervals. The results are as indicated in Table 5 and 6. Figures 3 and 5 represent the pie for +veDEV and -veDEV for generated and real data.

*Table 5:* Real Data Hourly interval deviation.
*Open value is the price at the beginning of interval, close is the price at the end of that interval. High is the highest attained price (uptrend) within the interval while low is lowest attained price (downtrend) within the interval. +veDEV is highest magnitude of price attained in uptrend converted into pipettes while -veDEV is lowest magnitude of price attained in downtrend converted into pipettes*

| Hr | OPEN | HIGH | LOW | CLOSE | +veDEV ((High-Open)*100000) | -veDEV ((Open-Low)*100000) |
|---|---|---|---|---|---|---|
| 1 | 1.08197 | 1.0827 | 1.0818 | 1.08217 | 73 | 17 |
| 2 | 1.0818 | 1.0823 | 1.0816 | 1.08198 | 50 | 20 |
| 3 | 1.08146 | 1.0823 | 1.08116 | 1.082 | 84 | 30 |
| . | . | . | . | . | . | . |
| . | . | . | . | . | . | . |
| . | . | . | . | . | . | . |
| 65 | 1.08029 | 1.08029 | 1.0792 | 1.07954 | 0 | 109 |
| 66 | 1.08033 | 1.08049 | 1.07988 | 1.08027 | 16 | 45 |

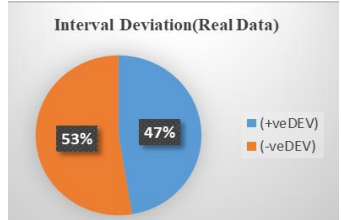

*Figure 4: Real Data hourly interval deviation.*
*Pie chart comparing +veDEV and -veDEV of Table 10*

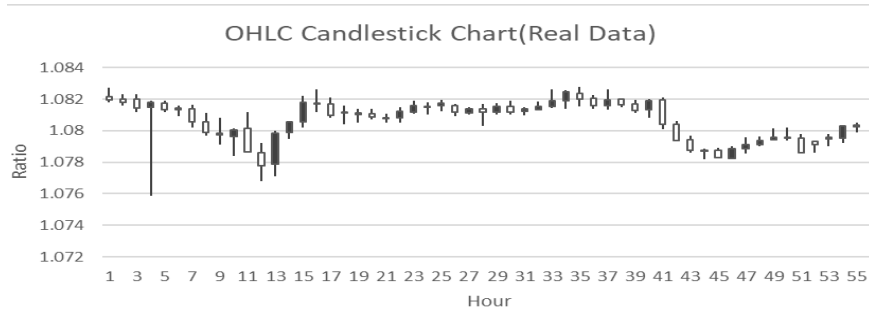

*Figure 5: Candlestick chart for 55 1 Hr intervals (Generated Data)*

*Table 6: Real Data % cumulative deviation; summary table of 10 batches*

| Batch No. | 1 | 2 | 3 | 4 | 5 | 6 | 7 | 8 | 9 | 10 |
|---|---|---|---|---|---|---|---|---|---|---|
| % +ve deviation | 47 | 56 | 45 | 56 | 45 | 50 | 42 | 50 | 40 | 47 |
| % -ve deviation | 53 | 44 | 55 | 44 | 55 | 50 | 58 | 50 | 60 | 53 |

## 3 Business case description

We first determine if essential characteristics identified by [22] exist in generated data to show relation with real in order to make inference on our experimental results. Visually, Figure 1 is comparable with commonly reported forex data graphs. Fluctuation is a sign of variability and it can also be observed that the data is non-linear and non-stationary. These features are essential characteristic in financial time series data [22]. The magnitude of price deviation on either of open price for regular intervals show that price oscillates uniformly on both sides. This behavior is manifest for real and simulated data and is an indicator of periodicity and seasonality in the data. When the results are evaluated and aggregate reward for *traders* compared with *intelligent market*, it has been demonstrated that *trader* lose in all cases i.e. 100% aggregate loss for *traders* in all simulations. This dynamic arises from the model's assumption that the *intelligent market* possesses complete information and the capacity to respond optimally following each trader's move. We can thus conclude that results from random trade type and risk appetite align with *trader* experience described in literature. It can thus be deduced that irrespective of the strategy used by retail traders in forex market, such intelligence is diffused and lost in the complexity of the trading environment. Even though different currency pairs manifest different interaction data, inherent behavior is shared across and the diversity can be accommodated with minor modifications in our algorithm.

From a business perspective, understanding the inherent risks in forex trading is essential. This should be viewed from the perspective of *traders*, financial institutions, and regulatory bodies. This study has highlighted the critical challenges that *traders* face in sustaining positive rewards within asymmetric information environments, where market structures may naturally favor entities with superior insight. For algorithmic traders, this calls for the development of advanced models and strategies optimized to recognize and adapt to such biases, enhancing both model training and real-time market analysis. Forex brokers and trading platforms have a major role in safeguarding fairness, while government regulatory bodies can implement robust oversight to protect individuals and communities vulnerable to the appeal of high-risk trading. By fostering an equitable market environment, these stakeholders can help mitigate exploitative practices and promote a sustainable trading ecosystem.

## 4 Conclusion

Based on market score and resultant interaction data, it can be inferred that if the market is analyzed from the premise of asymmetric information environment, more insights can be derived with potential of reducing forecasting efforts in forex trading. The procedure utilized is unique and confirms that forex trading is high risk if trivialized as simple binary problem. It is anticipated that further experiments to optimize market reward would result to more refined data for similarity tests and forex trading studies.